\documentclass[aps prapplied,twocolumn,superscriptaddress, longbibliography, floatfix]{revtex4-1}

\usepackage{graphicx}
\usepackage{amsmath}
\usepackage{bm}
\usepackage{color,soul}
\usepackage{gensymb}

\definecolor{mygreen}{rgb}{0.0, 0.7, 0.2}

\begin{document}
\setlength{\textfloatsep}{12pt}

\title{Efficient chip-based optical parametric oscillators from 590~nm to 1150~nm}


\author{Jordan R. Stone}
\email{jstone12@umd.edu}
\affiliation{Joint Quantum Institute, NIST/University of Maryland, College Park, MD 20742}
\affiliation{National Institute for Standards and Technology, Gaithersburg, MD 20899}

\author{Xiyuan Lu}
\affiliation{Joint Quantum Institute, NIST/University of Maryland, College Park, MD 20742}
\affiliation{National Institute for Standards and Technology, Gaithersburg, MD 20899}

\author{Gregory Moille}
\affiliation{Joint Quantum Institute, NIST/University of Maryland, College Park, MD 20742}
\affiliation{National Institute for Standards and Technology, Gaithersburg, MD 20899}

\author{Kartik Srinivasan}
\affiliation{Joint Quantum Institute, NIST/University of Maryland, College Park, MD 20742}
\affiliation{National Institute for Standards and Technology, Gaithersburg, MD 20899}


\date{\today}

\begin{abstract}
Optical parametric oscillators are widely used to generate coherent light at frequencies not accessible by conventional laser gain. However, chip-based parametric oscillators operating in the visible spectrum have suffered from pump-to-signal conversion efficiencies typically less than $0.1~\%$. Here, we demonstrate efficient optical parametric oscillators based on silicon nitride photonics that address frequencies between $260$ THz ($1150$ nm) and $510$ THz ($590$ nm). Pumping silicon nitride microrings near $385$ THz ($780$ nm) yields monochromatic signal and idler waves with unprecedented output powers in this wavelength range. We estimate on-chip output powers (separately for the signal and idler) between $1$~mW and $5$~mW and conversion efficiencies reaching $\approx15~\%$. Underlying this improved performance is our development of pulley waveguides for broadband near-critical coupling, which exploits a fundamental connection between the waveguide-resonator coupling rate and conversion efficiency. Finally, we find that mode competition reduces conversion efficiency at high pump powers, thereby constraining the maximum realizable output power. Our work proves that optical parametric oscillators built with integrated photonics can produce useful amounts of visible laser light with high efficiency.      
\end{abstract}


\maketitle


\section{\label{sec:intro}Introduction}

Lasers operating at visible and near-infrared (NIR) wavelengths are essential to modern science and technology \cite{bothwell2022resolving, awschalom2021development, matheus2019visible, weissleder2001clearer}, but affordable systems typically suffer from poor spectral purity and gaps in spectral coverage, while higher-performance options are large and expensive. The latter often rely on bulk nonlinear optics to spectrally translate longer-wavelength lasers to the targeted frequency, employing either sum-frequency or second-harmonic generation in $\chi^{(2)}$-nonlinear media \cite{tinsley2021watt, parrotta2013multiwatt, mildren2005discretely}. Their operational complexity and substantial power consumption (they often require liquid cooling systems) renders them impractical in many situations. Hence, it is desirable to transition the nonlinear wavelength conversion to a more scalable nonlinear optics platform, e.g., integrated photonics.

One approach is to leverage the wavelength flexibility inherent to optical parametric oscillators using nonlinear microresonators, which possess large optical quality factors ($Q$) and small mode volumes to intensify circulating light and promote efficient nonlinear interactions \cite{vahala2003optical, kippenberg2004kerr}. Recent studies of microresonator-based optical parametric oscillators ($\mu$OPOs) have demonstrated broad spectral separation between pump and generated light \cite{sayson2019octave, werner2012blue, jia2018continuous}, low-power operation \cite{lu2019milliwatt, lu2021ultralow}, and visible-wavelength access \cite{lu2020chip}. While both $\chi^{\rm{(2)}}$ and $\chi^{\rm{(3)}}$ $\mu$OPOs offer some wavelength flexibility, $\chi^{\rm{(3)}}$ systems are useful to generate visible light from a NIR pump; moreover, their natural availability in the popular silicon photonics platform~\cite{moss_new_2013} can enable their scalable fabrication and integration with other components, including pump lasers~\cite{park_heterogeneous_2020}. On the other hand, the reported or inferred (e.g., from optical spectra) conversion efficiencies are $\lesssim 0.1$\% \cite{sayson2019octave, lu2019milliwatt, lu2020chip, fujii2019octave, tang2020widely, domeneguetti2021parametric}, and the available output power is far too low for many applications (e.g., $<$10~$\mu$W for previous visible $\mu$OPOs~\cite{lu2020chip}). Realizing higher conversion efficiencies and output powers would enable a wide range of on-chip applications and broaden the reach of silicon photonics in the visible spectrum.   

Here, we demonstrate efficient $\mu$OPOs that generate coherent light within the spectral window between 260~THz and 510~THz (590~nm and 1150~nm). We measure conversion efficiencies between 3.5~\% and 14.5~\% with corresponding on-chip output powers greater than 1~mW (and as high as 5~mW). Our results spring from efficient broadband waveguide-resonator coupling, which we realize with pulley-waveguide geometries designed using coupled-mode simulations. In the rest of this paper, we first introduce the key $\mu$OPO physics and specify our experimental procedures. Then, we explain our coupled-mode simulations and present measurements to confirm their accuracy. Next, we present the optical spectra of 16 different $\mu$OPOs, from which we determine output powers and conversion efficiencies. Finally, we show how parasitic nonlinear processes currently constrain the maximum realizable output power. Our work is an important step forward in the quest for practical, chip-based sources of visible laser light using nonlinear optics. 

\begin{figure*}[ht]
    \centering
    \includegraphics[width=315 pt]{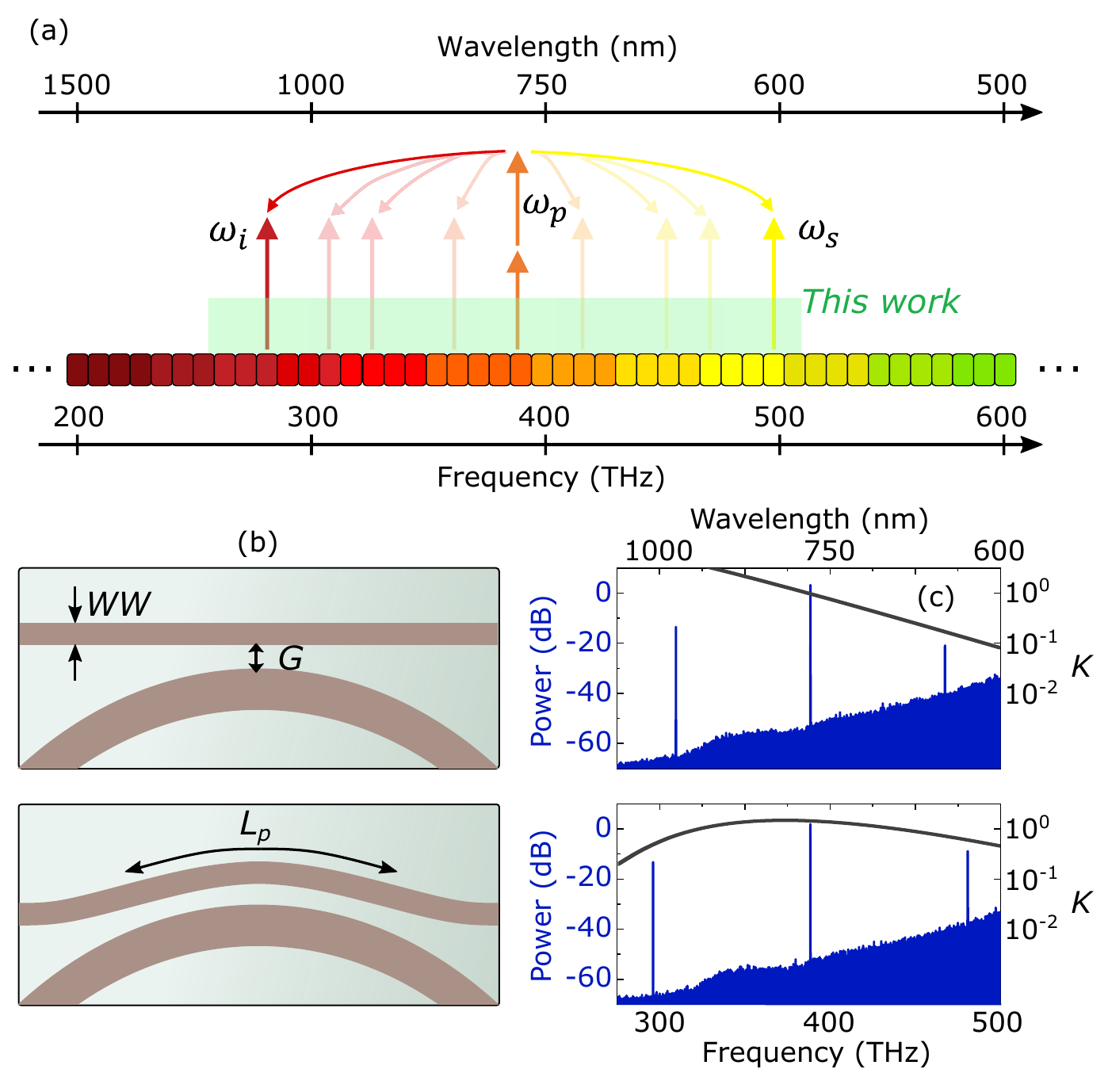}
    \caption{\textbf{Achieving high conversion efficiency ($CE$) across a wide spectral band with microresonator-based optical parametric oscillators ($\mu$OPOs).} (a) Conceptual depiction of $\chi^{(3)}$ optical parametric oscillation. Two pump photons (frequency $\omega_{\rm{p}}$) are converted to one signal ($\omega_{\rm{s}}$) photon and one idler ($\omega_{\rm{i}}$) photon. We focus on generating different $\mu$OPOs (faded arrows) within the spectral window between $260$~THz and $510$~THz, as marked by the green stripe. (b) Depictions of microring couplers, as viewed from above, which rely on evanescent coupling between the microring and either a straight access waveguide (top panel) or a pulley waveguide (bottom panel). Couplers are defined by three geometric parameters: the waveguide width, $WW$, the gap (or distance) between the microring and waveguide, $G$, and the length of waveguide, $L_{\rm{p}}$, that runs parallel to the microring. (c) Simulated coupling ratios, $K$ (gray), and measured $\mu$OPO spectra (blue) for identical microrings with different coupling geometries. The top panel data corresponds to a straight waveguide ($WW=375$ nm; $G=125$ nm) coupler, while bottom panel data corresponds to a pulley waveguide (same $WW$ and $G$; $L_{\rm{p}}=3$ $\mu$m).}
    \label{fig:one}
\end{figure*}

The $\mu$OPOs we consider generate monochromatic signal and idler waves from a continuous-wave (CW) pump laser through resonantly-enhanced degenerate four wave mixing (FWM), as depicted in Fig.~\ref{fig:one}a \cite{boyd2020nonlinear}. In experiments, we pump a fundamental transverse-electric (TE1) eigenmode of a silicon nitride microring near $385$ THz, and FWM transfers energy to TE1 signal and idler modes. In principle, the range of accessible output frequencies, as constrained only by energy conservation, is $DC$ to $2\omega_{\rm{p}}$, where $\omega_{\rm{p}}$ is the pump frequency. However, in practice this range is dictated by the group velocity dispersion (GVD), which must be engineered such that FWM to the targeted signal and idler modes is favored (simultaneously phase- and frequency-matched), but FWM to other modes is suppressed. In Appendix A, we recount our approach to dispersion engineering that is also described in Ref. \onlinecite{lu2020chip}.     

A separate challenge is to ensure that pump power is efficiently converted into output signal or idler power. Hence, we define the on-chip conversion efficiency as:

\begin{equation}\label{eq:CE}
CE = \frac{P_{\rm{s(i)}}}{P_{\rm{in}}},
\end{equation}
where $P_{\rm{s(i)}}$ is the signal (idler) power in the waveguide output, and $P_{\rm{in}}$ is the pump power in the waveguide input. Recent theoretical work has derived the maximum obtainable $CE$ as:

\begin{equation}\label{eq:CEmax}
CE_{\rm{s(i)}}^{\rm{max}}=\frac{1}{2}\frac{\kappa_{\rm{s(i)}}\kappa_{\rm{p}}}{\Gamma_{s(i)}\Gamma_{\rm{p}}}\frac{\omega_{\rm{s(i)}}}{\omega_{\rm{p}}},
\end{equation}
where $\kappa_{\rm{s(i)}}$ and $\kappa_{\rm{p}}$ are the waveguide-resonator coupling rates of the signal (idler) and pump modes, $\Gamma_{\rm{s(i)}}$ and $\Gamma_{\rm{p}}$ are the total loss rates (i.e., loaded linewidths) of the signal (idler) and pump modes, and $\omega_{\rm{s(i)}}$ and $\omega_{\rm{p}}$ are the frequencies of the signal (idler) and pump light, respectively \cite{sayson2019octave}. Clearly, obtaining large $CE$ involves engineering $\kappa$ for both the pump mode and targeted signal/idler modes. Such coupling engineering is a common problem in the nonlinear optics of Kerr microresonators; it arises, for example, in the efficient extraction of octave-spanning Kerr microcombs \cite{moille2019broadband}. The problem is that, given a straight waveguide evanescently coupled to a microring resonator, $\kappa$ decreases exponentially with frequency due to decreasing overlap of the microring mode with the waveguide mode. Hence, when the pump mode is critically coupled, the signal mode is undercoupled, resulting in low $CE$. Moreover, when $\omega_{\rm{s}}$ is a visible frequency, the smaller evanescent decay length compared to NIR frequencies exacerbates the challenge. One solution is to utilize so-called pulley waveguides, which increase the physical distance over which the waveguide and microring can exchange energy~\cite{moille2019broadband,shah_hosseini_systematic_2010}. Figure \ref{fig:one}b illustrates the physical differences between straight-waveguide couplers (top panel) and pulley-waveguide couplers (bottom panel), and it depicts the three geometric parameters that define such couplers in our study: The waveguide width, $WW$, the waveguide-resonator gap, $G$, and the pulley length, $L_{\rm{p}}$ (which approaches zero in the straight-waveguide limit). In Fig.~\ref{fig:one}c, we present measurements of $\mu$OPO spectra extracted from nominally identical microrings using either a straight-waveguide coupler (top panel) or a pulley-waveguide coupler (bottom panel). These measurements are representative of other comparisons between the two coupling schemes. While $P_{\rm{i}}$ is roughly the same in each case, $P_{\rm{s}}$ is approximately $20\times$ greater in the pulley waveguide. To explain the result, we show in the same panels the simulated coupling ratio, defined as $K=\kappa/\gamma$, where $\gamma=\Gamma-\kappa$ is the intrinsic loss rate. Near $\omega_{\rm{s}}$, $K$ is $\approx 8\times$ higher for the pulley-waveguide coupler.

\section{\label{sec:two}design and test of waveguide couplers}

\begin{figure}
    \centering
    \includegraphics[width=225 pt]{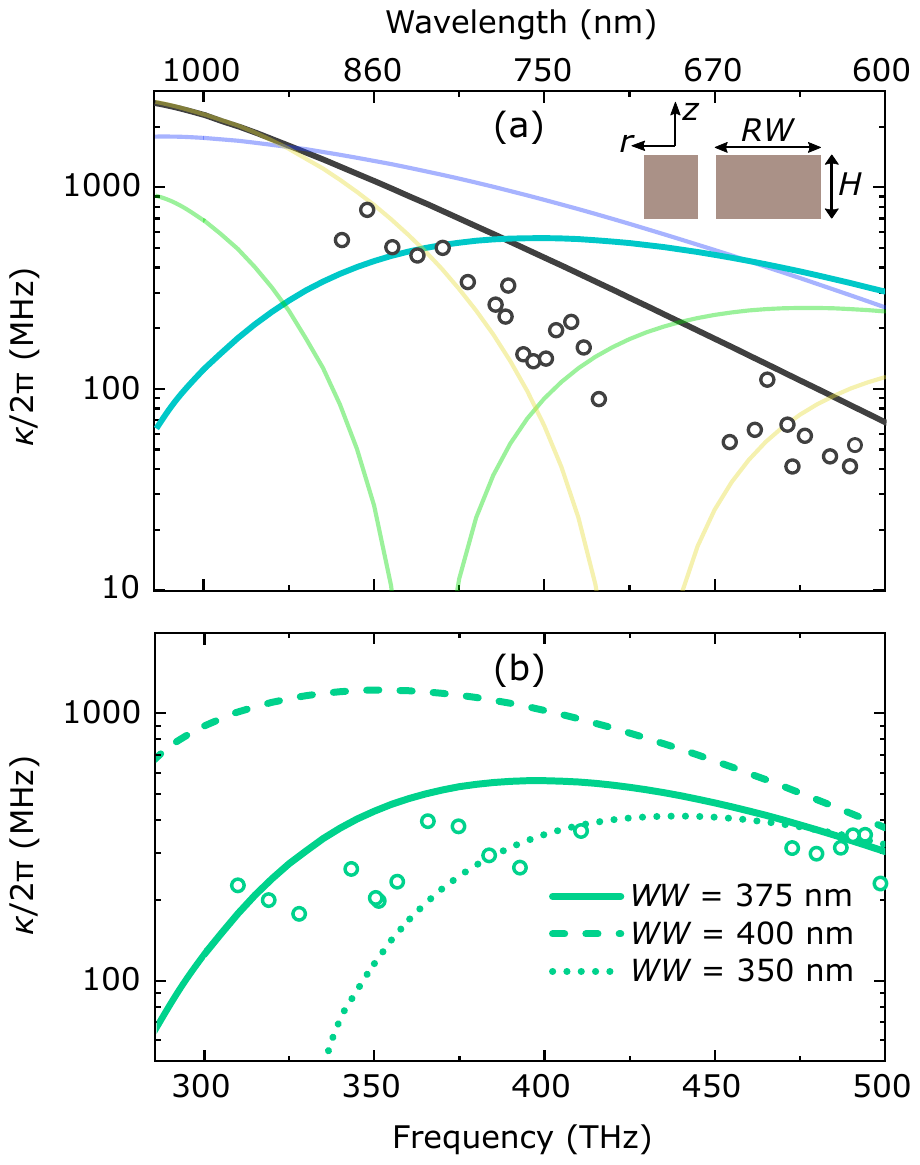}
    \caption{\textbf{Design, simulation, and measurement of coupling rate ($\kappa$).} (a) Simulated $\kappa$ spectra for a silicon nitride (SiN) microring with waveguide parameters $WW=375$ nm, $G=150$ nm, and $L_{\rm{p}}=0$ $\mu$m (bold gray curve), $2$ $\mu$m (faded blue), $3$ $\mu$m (bold turquoise), $4$ $\mu$m (faded green), and $5$ $\mu$m (faded yellow). The circular data points indicate measured $\kappa$ values for the coupler with $L_{\rm{p}}=0$ $\mu$m. The inset depicts the waveguide/microring cross section, including labels for the ring width ($RW$) and height ($H$), and shows axes for the coordinates $r$ and $z$ over which we integrate Eq. \ref{eq:coupling}. (b) Simulated $\kappa$ spectra for a SiN microring with waveguide parameters $G=150$ nm, $L_{\rm{p}}=3$ $\mu$m, and different values of $WW$. The circular data points indicate measured $\kappa$ values for the coupler with $WW=375$ nm. The uncertainty in the measured $\kappa$ values as determined from a nonlinear least squares fit to the resonator transmission spectrum is smaller than the data point size.}
    \label{fig:two}
\end{figure}

To design couplers for testing, we simulate $\kappa$ spectra for a variety of coupling geometries with the goal of achieving $K\approx1$ (i.e., critical coupling) at frequencies between $260$ THz and $510$ THz. Notably, achieving high $CE$ only requires that $\kappa$ be optimized at $\omega_{\rm{p}}$ and either $\omega_{\rm{s}}$ or $\omega_{\rm{i}}$, depending on which output wave (signal or idler) is to be used. (Moreover, ideally, the unused wave is undercoupled to reduce threshold power). At the same time, broadband near-critical coupling is preferable, since then a single coupling geometry is robust against design imperfections, and it may be used for many different $\mu$OPOs. Our simulations are based on a coupled mode theory for optical waveguides \cite{huang1994coupled}, which calculates $\kappa$ according to: 

\begin{equation} \label{eq:kappa}
\kappa = \frac{c}{2\pi R n_{\rm{g}}}|k_{\rm{t}}|^2,
\end{equation}
where $R$ is the microring outer radius, $c$ is the speed of light, $n_{\rm{g}}$ is the group refractive index, and $k_{\rm{t}}$ is a coupling coefficient defined as:
 
 \begin{equation} \label{eq:coupling}
 k_{\rm{t}} = \frac{i\omega}{4}\int_{L}\left [\int_{A}(\epsilon_{\rm{WG}}-\epsilon_{\rm{R}})\boldsymbol{E_{\rm{R}}^*}\cdot\boldsymbol{E_{\rm{WG}}}\rm{d}\it{r}\rm{d}\it{z} \right ] e^{i\phi}\rm{d}\it{l},
 \end{equation}
where $\epsilon_{\rm{WG(R)}}$ is the dielectric permittivity of the access waveguide (microring), $\boldsymbol{E_{\rm{WG(R)}}}$ is the electric field of the waveguide (microring) eigenmode, and $\phi$ is an accumulated phase accounting for the difference in the waveguide and microring propagation constants. The coordinates $r$ and $z$ are horizontal and vertical coordinates in-plane with the microring/waveguide cross section, as labeled in the Fig. \ref{fig:two}a inset, and $l$ follows the direction of light propagation. The central integral in Eq.~(\ref{eq:coupling}) evaluates the evanescent overlap between the microring and waveguide modes at the frequency $\omega$. For more details, see Ref. \cite{moille2019broadband}. 
\begin{figure*}[ht]
    \centering
    \includegraphics[width=500 pt]{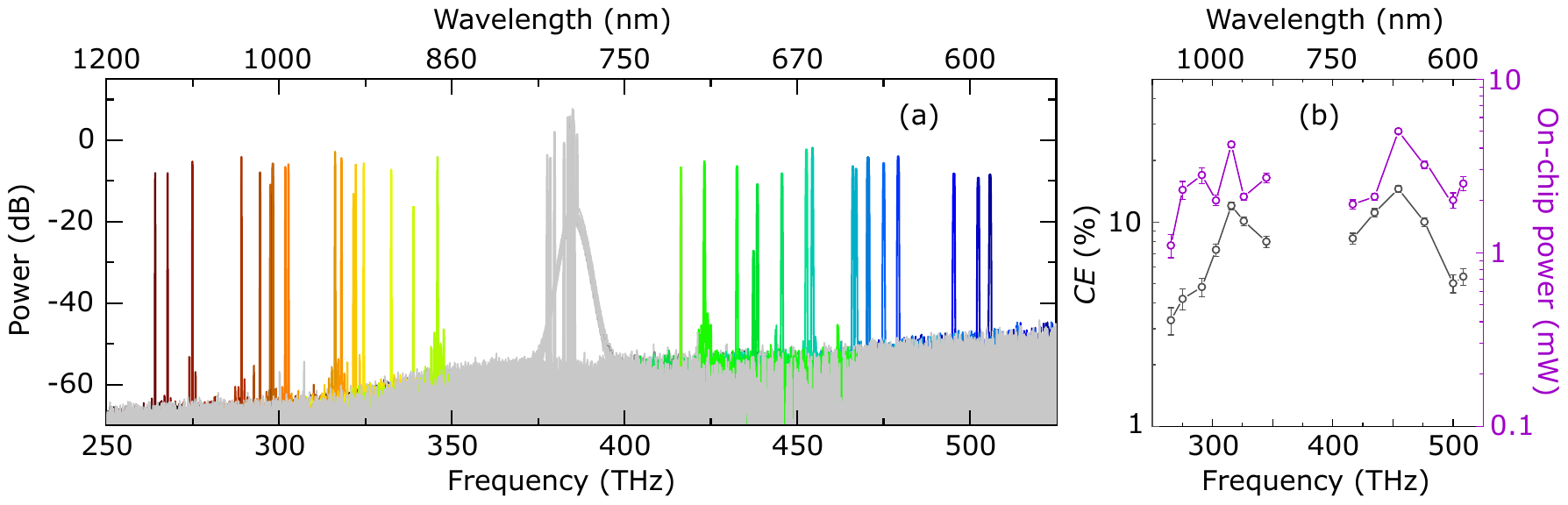}
    \caption{\textbf{$\mu$OPO spectra, conversion efficiency, and output power.} (a) $\mu$OPO spectra from a series of devices recorded with an optical spectrum analyzer (OSA). Here, the y axis is normalized such that $0$ dB is equal to $1$ mW. No correction factors (e.g., from optical losses) are applied to these data. The transmitted pump light is shown in gray, and the signal/idler light is shown in color. (b) $CE$ (gray) and $P_{\rm{i}}$ or $P_{\rm{s}}$ (purple; left and right portions of data, respectively) versus frequency, taking into account optical losses between the waveguide and OSA (see Appendix B). Error bars are estimated from the precision uncertainties in our measurements of optical power.}
    \label{fig:three}
\end{figure*}
Figure~\ref{fig:two}a shows simulated $\kappa$ spectra for a SiN microring with nominal $R=25$ $\mu$m, ring width $RW=825$ nm, and height $H=500$ nm, which are chosen to suitably engineer the GVD (see Appendix A). In addition, we choose $WW=375$ nm, $G=150$ nm, and $L_{\rm{p}}$ from $0$ $\mu$m (i.e., a straight-waveguide coupler) to $5$ $\mu$m. Increasing $L_{\rm{p}}$ results in larger $\kappa$ at higher frequencies compared to the straight-waveguide coupler (dark grey line), but it introduces resonances in the $\kappa$ spectra (i.e., regions where $\kappa\rightarrow0$) that arise from the coherent energy exchange between the microring and waveguide. These resonances blueshift when $L_{\rm{p}}$ is increased. Based on these data, we select $L_{\rm{p}}=3$ $\mu$m for further study because it minimizes variations in $\kappa$ over the targeted spectral region. 

Next, we optimize $WW$ and $G$. The predominant effect of changing $G$ is to vertically shift the entire $\kappa$ spectrum; i.e., $G$ has little impact on the spectral location of the coupling resonances. However, the relationship between $\kappa$ and $WW$ is more complex. Within the range of values considered, larger $WW$ increases $\kappa$ because the waveguide propagation constant shifts closer to the microring propagation constant, and the evanescent overlap between the microring and waveguide modes is not appreciably changed. At the same time, coupling resonances are redshifted. Figure \ref{fig:two}b shows $\kappa$ spectra for $G=150$ nm, $L_{\rm{p}}=3$ $\mu$m, and three values of $WW$. Apparently, the flattest $\kappa$ spectra are realized for $WW$ between $375$ and $400$ nm. After choosing $WW$, $G$ may be chosen to realize critical coupling near $\omega_{\rm{p}}$.

To assess the accuracy of our simulations, we fabricate an array of SiN microrings with systematic coupling parameter variations (see Appendix C for details on our fabrication process), and we measure $\kappa$ for each device. Specifically, we use either $L_{\rm{p}}=0$ or $L_{\rm{p}}=3$ $\mu$m, $G$ between $110$~nm and $160$~nm, and $WW$ between $350$~nm and $400$~nm. To measure $\kappa$, we carry out mode spectroscopy using a CW Titanium Sapphire (TiS) laser, which is tunable from $305$~THz to $415$~THz ($720$~nm to $980$~nm). In addition, we can perform sum frequency generation with the TiS laser and a $154$~THz ($1950$~nm) laser to generate coherent light from $460$~THz to $510$~THz ($590$~nm to $650$~nm). We find that simulations predict slightly larger $\kappa$ values than we measure; to compensate, we reduce $G$ by approximately $20$~nm in experiments. In Fig.~\ref{fig:two}a, we present $\kappa$ measurements (gray circles) of a straight-waveguide-coupled device with $WW=375$ nm and $G=125$ nm. The measured $\kappa$ values are slightly lower than the corresponding simulation with $G=150$ nm, but both data decrease exponentially with frequency, which is a known characteristic of straight-waveguide couplers~\cite{moille2019broadband}. Specifically, we measure $\kappa \approx 800$ MHz near $350$~THz, but it drops sharply to $\kappa \approx 40$ MHz near $500$ THz. In contrast, we observe a more achromatic $\kappa$ spectrum in a pulley-waveguide-coupled device with $WW=375$ nm, $G=135$ nm, and $L_{\rm{p}}=3$ $\mu$m. Our measurements (green circles) are shown in Fig. \ref{fig:two}b, and they agree with the corresponding simulation with $G=150$ nm. Our measurements indicate that, between $300$~THz and $500$~THz, $\kappa$ takes on values over the relatively narrow range (compared to the straight-waveguide coupler) of $180$~MHz to $400$~MHz. We also measure $\gamma \approx 300$ MHz that is approximately independent of optical frequency. Therefore, according to Eq. \ref{eq:CEmax}, our best pulley-waveguide-coupled devices should support many different $\mu$OPOs with $CE>1$\%.   

\section{\label{sec:three}$\mu$OPO generation and conversion efficiency measurements}

To test our prediction, we record $\mu$OPO spectra with a calibrated optical spectrum analyzer (OSA) and calculate $P_{\rm{s}}$, $P_{\rm{i}}$, and $CE$ values after accounting for optical losses between the waveguide and OSA (for details, see Appendix B). To generate $\mu$OPOs, we tune $\omega_{\rm{p}}$ into resonance, starting blue detuned and decreasing $\omega_{\rm{p}}$ until $P_{\rm{s}}$ and $P_{\rm{i}}$ are maximized. We repeat this procedure for different $P_{\rm{in}}$ values with the goal of maximizing $CE$. To ensure a variety of $\omega_{\rm{s(i)}}$ values, we engineer the GVD by systematically varying $RW$ in different devices (see Ref. \cite{lu2020chip} and Appendix A). We utilize pulley-waveguide couplers such as those characterized in Fig. \ref{fig:two}, and we find that $CE$ is maximized for $L_{\rm{p}}=3$ $\mu$m, $WW$ between $375$~nm and $400$~nm, and $G$ between $125$~nm and $135$~nm. In Fig. \ref{fig:three}a, we present a compiled set of $\mu$OPO spectra that we extract from pulley-waveguide couplers with parameters in the above optimum range. In most cases, both $P_{\rm{s}}$ and $P_{\rm{i}}$ are greater than $1$ mW, and $P_{\rm{in}}$ is typically between $30$~mW and $45$~mW. However, there are atypical $\mu$OPO spectra for which either $P_{\rm{s}}$ or $P_{\rm{i}}$ is $\ll1$~mW. Most likely, these result from mode interactions that locally alter the microring GVD and $Q$ \cite{herr2014mode,ramelow2014strong}. Table II in Appendix B lists the individual pump, signal, and idler frequencies and powers for each $\mu$OPO spectrum shown in Fig.~\ref{fig:three}(a).

To characterize the $\mu$OPO performance, we calculate from Fig. \ref{fig:three}a the largest values of $P_{\rm{s}}$ and $P_{\rm{i}}$ in spectral bins spanning approximately $20$ THz each, and we plot the results in Fig. \ref{fig:three}b along with the corresponding $CE$ values. We find $P_{\rm{i}}>1$ mW from $264$~THz to $346$~THz and $P_{\rm{i}}>2$ mW from $275$~THz to $346$~THz. In the best case, we generate $4$ mW of idler power at $315$~THz using $P_{\rm{in}}=34$ mW, which equates to $CE\approx 12$\%. Meanwhile, $P_{\rm{s}}>1.9$ mW from $416$~THz to $506$~THz, and in the best case, we generate $5$ mW of signal power at $454$~THz using $P_{\rm{in}}=34$ mW, which equates to $CE\approx 14.5$\%. Moreover, as expected from our simulations and measurements of $\kappa$ and Eq. \ref{eq:CEmax}, $CE$ decreases in the spectral wings as a result of smaller $\kappa$. 

\begin{figure}[t]
    \centering
    \includegraphics[width=225 pt]{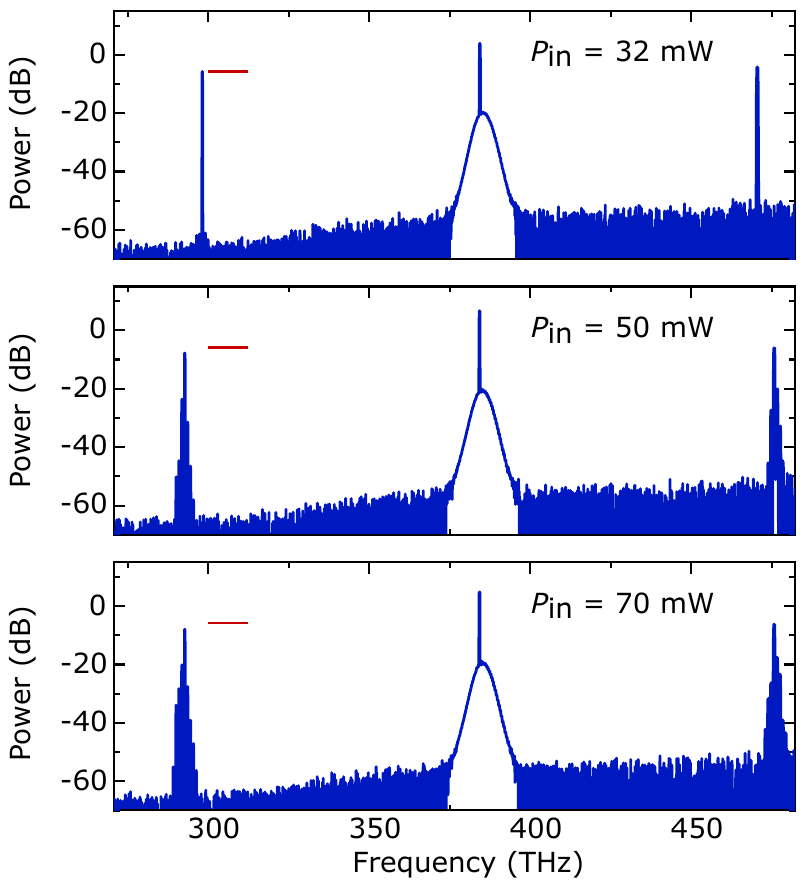}
    \caption{\textbf{Mode competition constrains the $\mu$OPO output power.} The figure depicts $\mu$OPO spectra at three different pump powers, $P_{\rm{in}}$. For $P_{\rm{in}}=32$ mW (top panel), only the pump, signal, and idler modes oscillate. As $P_{\rm{in}}$ is increased (middle and bottom panels), mode switching occurs, and mode competitions distribute energy to modes other than the targeted signal/idler pair, leading to reduced $P_{\rm{s(i)}}$. To guide the eye, we include a red line in each panel that marks the top-panel $P_{\rm{i}}$ level.}
    \label{fig:four}
\end{figure}

\section{\label{sec:four}Limitations on output power}

Finally, we discuss the limits to $\mu$OPO output power and analyze an example. The relationship between $P_{\rm{s(i)}}$, $P_{\rm{in}}$, and other experimental parameters has been analyzed theoretically \cite{stone2022conversion}. Therein, it was predicted that $P_{\rm{s(i)}}$ increases with $P_{\rm{in}}$ for $P_{\rm{in}} \gtrsim P_{\rm{th}}$, but further increases in $P_{\rm{in}}$ lead to saturation or even reduction of $P_{\rm{s(i)}}$. The reason is that parasitic FWM processes compete with the targeted $\mu$OPO process. The predominant parasitic FWM process that we observe in experiments is mode competition between the targeted signal/idler modes and their spectral neighbors \cite{stone2022conversion}. In Fig. \ref{fig:four}, we show the $\mu$OPO spectrum for a single device as $P_{\rm{in}}$ is increased. For $P_{\rm{in}}=32$ mW, only the pump, signal, and idler modes oscillate, and $P_{\rm{s(i)}}>1$ mW (top panel). The idler power in this case is marked by the red line in each panel. When $P_{\rm{in}}=50$ mW (middle panel), $\omega_{\rm{s}}$ and $\omega_{\rm{i}}$ shift to higher and lower frequencies, respectively. This behavior was predicted in Ref. \cite{stone2022conversion} and termed `mode switching.' In addition, other modes with frequencies close to $\omega_{\rm{s(i)}}$ begin to oscillate and steal energy from the $\mu$OPO. Hence, $P_{\rm{i}}$ decreases to a level below the red line, despite the increase in $P_{\rm{in}}$. When $P_{\rm{in}}$ is further increased to $70$ mW (bottom panel), $P_{\rm{s(i)}}$ remains approximately the same, but the power in the competing modes increases. The behavior demonstrated in this example is ubiquitous within our $\mu$OPO devices and explains why $P_{\rm{s(i)}}$ cannot be increased arbitrarily by increasing $P_{\rm{in}}$. Still, increasing $CE$ and $P_{\rm{s(i)}}$ beyond the levels we demonstrate may be possible using alternate phase- and frequency-matching strategies, such as that reported in Ref. \cite{zhou2022hybrid}. As it stands, the achieved power levels are relevant for some applications, such as spectroscopy of various coherent near-infrared and visible systems \cite{xu2007coherent, brazhnikov2020two}. 

\section{\label{sec:five}Discussion}

In conclusion, we have demonstrated energy-efficient $\mu$OPOs with practically-relevant output powers in a crucial portion of the visible and near-infrared spectrum. Our results are enabled by broadband pulley-waveguide couplers that we design using coupled-mode simulations. For the most widely-separated $\mu$OPOs, we observe relatively lower $CE$ values that are consistent with undercoupled signal and idler modes. Hence, an important focus for future work is to extend the spectral bandwidth over which devices are nearly critically coupled, thus broadening the range of frequencies that can be efficiently extracted into a single waveguide. Other possible approaches include using frequency-specific coupling geometries (e.g., one may optimize $\kappa$ at the $\mu$OPO-specific frequencies $\omega_{\rm{s}}$ and $\omega_{\rm{p}}$, while neglecting the idler), or to couple multiple waveguides - each designed for different portions of the $\mu$OPO spectrum - to one microring. Moreover, new strategies should be devised to avoid parasitic FWM processes and increase the realizable output power. Nonetheless, our work is a compelling demonstration that $\mu$OPOs can help satisfy the demand for compact sources of visible laser light. 

\begin{acknowledgements}
We thank Khoi Hoang and Tahmid Sami Rahman for carefully reading the manuscript and giving useful feedback. This project is funded by the DARPA LUMOS and NIST-on-a-chip programs.  
\end{acknowledgements}

\appendix
\section{Engineering GVD to control the $\mu$OPO spectrum}
\begin{figure*}[ht]
    \centering
    \includegraphics[width=450 pt]{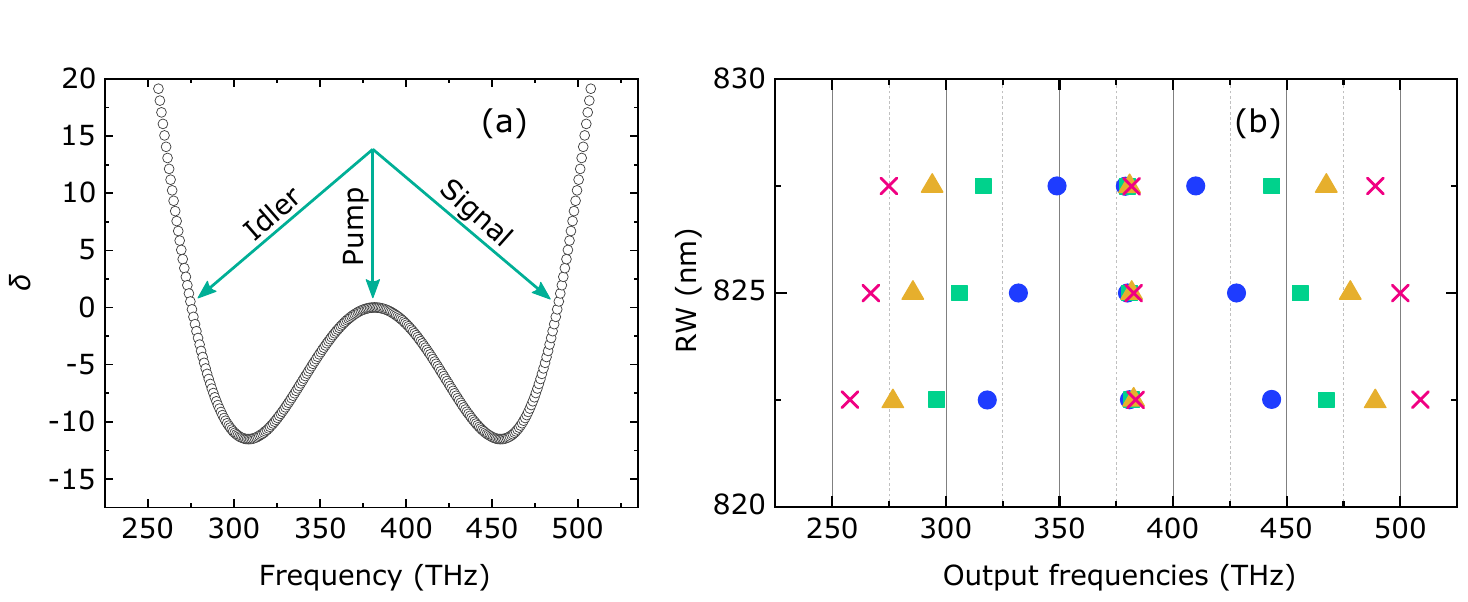}
    \caption{\textbf{Engineering dispersion in $\mu$OPO systems.} (a) Simulated frequency mismatch parameter, $\delta$, versus frequency. Arrows point to the generated frequencies, $\omega_{\rm{s}}$ and $\omega_{\rm{i}}$, and the pump frequency, $\omega_{\rm{p}}$. (b) Predicted $\mu$OPO frequencies for different values of $RW$ and $\omega_{\rm{p}}$. Different colors/shapes correspond to different pump modes, with $\omega_{\rm{p}}$ (near $385$~THz) increasing from the blue circle to the red cross.}
    \label{fig:a1}
\end{figure*}
The microresonator GVD determines which frequencies, if any, are phase- and frequency-matched for FWM. When designing $\chi^{\rm{(3)}}$ $\mu$OPOs, a convenient quantity related to GVD is the frequency mismatch parameter, $\delta(\nu)=\nu_{\rm{s}}+\nu_{\rm{i}}-2\nu_{\rm{p}}$, where we use a slightly different notation than in the main text; namely, $\nu_{\rm{s(i)}}$ are the set of possible signal (idler) mode frequencies, and $\nu_{\rm{p}}$ is the pump mode frequency. Degenerate FWM is allowed when $\delta(\nu) \gtrsim 0$ (small positive $\delta$ is required to compensate nonlinear shifts). Figure \ref{fig:a1}a shows a typical $\delta(\nu)$ spectrum as engineered for a $\chi^{\rm{(3)}}$ $\mu$OPO system. The spectrum exhibits normal dispersion around $\nu_{\rm{p}}$, which is necessary to suppress comb formation \cite{lu2020chip}, and higher-order GVD causes $\delta$ to turn and pass through zero \cite{stone2022conversion}. In general, oscillation occurs on the mode pair with smallest positive $\delta$.

The $\delta$ spectrum is controlled through the microring geometry; namely, the $RW$ parameter. Moreover, a given microring may support several different $\mu$OPOs, depending on $\omega_{\rm{p}}$. In our devices, a group of four or five pump modes will yield different $\mu$OPO spectra, with larger $\omega_{\rm{p}}$ increasing the signal-idler frequency separation. In Fig. \ref{fig:a1}b, we mark $\omega_{\rm{p}}$, $\omega_{\rm{s}}$, and $\omega_{\rm{i}}$ for four different $\mu$OPOs at each of three $RW$ values. Importantly, fine-tuning $RW$ (combined with a strategic choice of $\omega_{\rm{p}}$) can yield $\mu$OPOs with generated frequencies anywhere in the targeted spectral window. 

\section{Estimating on-chip powers from OSA measurements}
Future integrated-photonics systems will route laser light from source to destination (e.g., from a $\mu$OPO-generating microring to an integrated atomic vapor cell) via low-loss optical waveguides. Hence, we estimate on-chip powers and conversion efficiencies, which requires a careful calibration of optical losses between the waveguide and OSA. In general, such losses are dispersive, so we measure them for each device at a comprehensive set of optical frequencies. We use lensed fibers to in- and out-couple light to and from the waveguide. The lensed-fiber output is split using a fiber-based $90/10$ coupler that is designed for a wide wavelength span around $780$ nm. The `$90$' output port is connected to the OSA via optical fiber, while light from the other port is connected to a photodetector. To estimate losses, we use a power meter to record the optical power at various points in the experiment; specifically, we take measurements at the lensed-fiber input, lensed-fiber output, and OSA input. Finally, we note any discrepancies between the OSA input and OSA reading. Table \ref{tab:1} shows an example taken at $306$~THz ($980$~nm) for one $\mu$OPO device. Adding the different losses yields the OSA-to-on-chip power conversion factor, which allows us to convert $\mu$OPO spectra to on-chip signal/idler powers. Moreover, knowledge of the waveguide-to-lensed-fiber coupling losses allows us to estimate the on-chip pump power in experiments, which we use to estimate $CE$ values.     
\begin{widetext}
\begin{center}
\begin{table}[h!]
 \begin{tabular}{||c c c||} 
 \hline
 Measurement point & Power ($\mu$W) & Loss (dB) \\ [0.5ex] 
 \hline\hline
 Input to waveguide & 10 & N/A \\
 \hline
 Output from waveguide & 2.2 & 3.3 per waveguide facet \\
 \hline
 Input to OSA (after coupler) & 1.4 & 2 \\
 \hline
 OSA reading & 0.7 & 3 \\
 \hline
 \textbf{Total} & & \textbf{8.3} \\
 \hline
\end{tabular}
\caption{Example of loss calibrations at $306$ THz ($980$ nm) to determine the on-chip power conversion factor.}
\label{tab:1}
\end{table}
\end{center}
\end{widetext}

Finally, we tabulate the signal and idler frequencies and on-chip powers for each of the $16$ spectra shown in Fig. \ref{fig:three}a. These data are presented in 
\begin{widetext}
\begin{center}
\begin{table}[h!]
 \begin{tabular}{||c c |c c | c c ||} 
 \hline
 $\omega_{\rm{p}}/2\pi$ (THz) & $P_{\rm{in}}$ (mW) & $\omega_{\rm{s}}/2\pi$ (THz) & $P_{\rm{s}}$ (mW) & $\omega_{\rm{i}}/2\pi$ (THz) & $P_{\rm{i}}$ (mW) \\ [0.5ex] 
 \hline\hline
 385 & 34 & 423 & 2.2 & 346 & 2.7 \\
 \hline
 378 & 26 & 416 & 1.8 & 339 & 0.2 \\
 \hline
 378 & 20 & 432 & 1.8 & 324 & 2.1 \\
  \hline
 380 & 25 & 437 & 0.02 & 322 & 1.8 \\
  \hline
 385 & 27 & 438 & 0.8 & 332 & 1.4 \\
  \hline
 384 & 30 & 446 & 1.3 & 321 & 0.4 \\
  \hline
 385 & 36 & 454 & 5.0 & 318 & 2.6 \\
  \hline
 385 & 35 & 454 & 4.4 & 316 & 4.0 \\
  \hline
 384 & 28 & 466 & 1.9 & 303 & 1.8 \\
  \hline
 384 & 46 & 467 & 1.7 & 302 & 1.6 \\
  \hline
 384 & 32 & 470 & 3.3 & 298 & 1.9 \\
  \hline
 386 & 33 & 475 & 2.2 & 297 & 0.6 \\
  \hline
 384 & 59 & 479 & 3.4 & 289 & 2.7 \\
  \hline
 385 & 55 & 495 & 2.4 & 275 & 2.3 \\
  \hline
 385 & 34 & 502 & 2.0 & 268 & 1.1 \\
  \hline
 385 & 47 & 506 & 2.5 & 264 & 1.1 \\
 \hline
\end{tabular}
\caption{Pump, signal, and idler frequencies and on-chip powers for the $16$ $\mu$OPO spectra presented in Fig. \ref{fig:three}a.}
\label{tab:1}
\end{table}
\end{center}
\end{widetext}

\section{Fabrication methods}
To create device layouts, we use the Nanolithography Toolbox, a free software package developed by NIST \cite{balram2016nanolithography}. We deposit stoichiometric SiN (Si$_3$N$_4$) by low-pressure chemical vapor deposition on top of a $3$ $\mu$m-thick layer of SiO$_{\rm{2}}$ on a $100$ mm diameter Si wafer. We fit ellipsometer measurements of the wavelength-dependent SiN refractive index and layer thicknesses to an extended Sellmeier model. The device pattern is created in positive-tone resist by electron-beam lithography and then transferred to SiN by reactive ion etching using a CF$_{\rm{4}}$/CHF$_{\rm{3}}$ chemistry. After cleaning the devices, we anneal them for four hours at $1100$\degree C in N$_{\rm{2}}$. Next, we perform oxide lift-off to make devices air-clad around their left, right, and top sides. The facets of the chip are then polished for lensed-fiber coupling. After polishing, the chip is annealed again.
\bibliography{Bibliography}

\end{document}